\documentclass[%
 prl,
 amsmath,amssymb,
 twocolumn,
author-year,
superscriptaddress,
]{revtex4-2}

\usepackage{graphicx,dcolumn,amsmath,amssymb,mathrsfs,titlesec,hyperref,enumerate,bm,subfigure,float,comment,mathrsfs,xcolor}
\definecolor{linkColor}{rgb}{1,0,0}
\definecolor{citeColor}{rgb}{1,0,0}
\hypersetup{pdfborder={0 0 0}, colorlinks=true,urlcolor=linkColor,citecolor=citeColor}

\setcitestyle{authoryear,round}
\usepackage{totcount}
\usepackage{xcolor}
\usepackage{soul}
\usepackage{ulem}
\usepackage{lipsum}
\usepackage{lmodern}
\usepackage{tcolorbox}

\newtotcounter{citnum} 
\def\oldbibitem{} \let\oldbibitem=\bibitem
\def\bibitem{\stepcounter{citnum}\oldbibitem}

\begin{document}

\title{Deep Learning Segmentation of Complex Features in Atomic-Resolution Phase Contrast Transmission Electron Microscopy Images}

\author{Robbie Sadre}
\email{rssadre@lbl.gov}
\affiliation{Computational Research Division, Lawrence Berkeley National Laboratory, 1 Cyclotron Road, Berkeley, CA 94720}

\author{Colin Ophus}
\email{cophus@gmail.com}
\affiliation{NCEM, Molecular Foundry, Lawrence Berkeley National Laboratory, 1 Cyclotron Road, Berkeley, CA 94720}

\author{Anstasiia Butko}
\affiliation{Computational Research Division, Lawrence Berkeley National Laboratory, 1 Cyclotron Road, Berkeley, CA 94720}

\author{Gunther H Weber}
\affiliation{Computational Research Division, Lawrence Berkeley National Laboratory, 1 Cyclotron Road, Berkeley, CA 94720}

\begin{abstract}
Phase contrast transmission electron microscopy (TEM) is a powerful tool for imaging the local atomic structure of materials. TEM has been used heavily in studies of defect structures of 2D materials such as monolayer graphene due to its high dose efficiency. However, phase contrast imaging can produce complex nonlinear contrast, even for weakly-scattering samples. It is therefore difficult to develop fully-automated analysis routines for phase contrast TEM studies using conventional image processing tools. For automated analysis of large sample regions of graphene, one of the key problems is segmentation between the structure of interest and unwanted structures such as surface contaminant layers. In this study, we compare the performance of a conventional Bragg filtering method to  a deep learning routine based on the U-Net architecture. We show that the deep learning method is more general, simpler to apply in practice, and produces more accurate and robust results than the conventional algorithm. We provide easily-adaptable source code for all results in this paper, and discuss potential applications for deep learning in fully-automated TEM image analysis.


\end{abstract}

\maketitle

\section{Introduction}

High resolution transmission electron microscopy (HRTEM) is a very powerful technique for imaging atomic structure, due to its extremely high spatial resolution. HRTEM has found wide application in studies of the local atomic structure of two-dimensional materials, such as graphene \citep{meyer2007structure, warner2009structural, meyer2011experimental, mas20112d, robertson2013atomic, rasool2013measurement, rasool2015atomic}.Monolayer graphene is composed of a single 2D sheet of carbon atoms, with the same in-plane structure as the parent material graphite \citep{cooper2012experimental}. Most synthesis methods that can produce monolayer graphene will also produce defect structures, including point defects \citep{hashimoto2004direct, jeong2008stability,  kotakoski2014imaging}, edges \citep{russo2012atom, wang2014direct}, and line defects, such as grain boundaries \citep{huang2011grains, yu2011control}. 

Grain boundaries in graphene are scientifically interesting because of their distinctive mechanical \citep{grantab2010anomalous, rasool2013measurement, lee2013high}, electronic \citep{jauregui2011electronic, tapaszto2012mapping, fei2013electronic}, optical \citep{duong2012probing, podila2012effects}, and chemical properties \citep{yasaei2014chemical, kim2014selective}. In a previous study, \cite{ophus2015large} used experimental HRTEM imaging and numerical simulations to map out the parameter space of single-layer graphene grain boundaries, as a function of misorientation and boundary tilt angle. This previous work used semi-automated analysis routines to map out the atomic positions of the boundaries. Once boundary regions were identified, the atomic position analysis was almost entirely automated. However, each of these boundaries had to be hand selected and individually masked, due to the presence of surface contaminants. These contaminants are likely amorphous carbon \citep{zhang2019large}, which tends to be attracted by the charging induced by the electron beam to the boundary regions. This previous work did not utilize a reliable fully-automated computational method for segmenting between the desirable and undesirable atomic structures.


Recently however, new image analysis methods have been developed under the umbrella of deep learning \citep{garcia2017review}. Deep learning as an approach to data processing problems has substantially grown in popularity over the last decade. This can be attributed to increasing availability of large labeled datasets, such as Image-Net \citep{imagenet_cvpr09}, breakthrough research publications in the field \citep{krizhevsky2010imagenet}, and availability of high performance deep learning frameworks, such as PyTorch \citep{paszke2019pytorch} and TensorFlow \citep{tensorflow2015-whitepaper}. Convolutional Neural Networks (CNNs) have been used for various different image processing tasks, such as image classification (recognition of the object class within an image), object detection (classification and detection of objects in an image as well as generation of the bounding box around the object) and semantic segmentation (pixel-wise classification of an image).


Various works \citep{DL_defect2, DL_Defect1} have successfully applied deep learning methodologies to analyzing atomic defects in microscopy images of materials. 
In particular, \cite{madsen2018deep} used a deep learning network trained on simulated TEM data to recognize local structures in graphene images.

Various studies have made use of neural networks for segmentation of images of cells, such as \cite{akram2016cell,al2018deep}, as well as other biological datasets, such as vasculature stacks \citep{teikari2016deep}, brain tumors \citep{dong2017automatic}, and neuron structures \citep{dahmen2019sparse}. Many works have introduced application specific architectures for their studies, e.g., \cite{kassim2017deep, roberts2019defectnet}.

For the segmentation task considered in this paper, we utilize the U-Net architecture as described in\citep{unet}, due to its proven ability to achieve high performance results on image segmentation tasks with limited training data. This aspect is crucial, as large databases of labeled data are typically not readily available for most scientific imaging applications. U-Net has been applied to various datasets, such as urine microscopy images \citep{aziz2018improved}, ADF-STEM images \citep{ge2018deep}, corneal endothelial cell images \citep{daniel2019automated}, and fluorescently labelled cell nuclei images \citep{gudla2019deep}. 
Many other works performed similar microscopy segmentation tasks on the nanoscale using modified versions of the U-Net Architecture such as EM-Net \citep{khadangi2020net}, Fully Residual U-Net \citep{gomez2019deep}, Inception U-Net \citep{punn2020inception}, and the domain adaptive approach with two coupled U-Nets \citep{bermudez2018domain}.
In this paper, we develop a deep learning-based image segmentation pipeline for detecting surface contaminants in HRTEM images of graphene and compare it to a
conventional Bragg filtering approach \citep{hytch1997geometric, galindo2007peak}. Section~\ref{sec:materail}
reviews materials and methods used in our study. First, we describe
image acquisition and preprocessing methodologies as well as labeling training and
test data for our modeling approach. We also review Bragg filtering as a classical
image segmentation approach for detecting surface contaminants in graphene,
which serves as a baseline model. We then introduce our new method that trains and 
evaluates a U-Net based neural network architecture using k-fold cross validation.
We demonstrate that our neural network's automated feature learning capabilities
outperform Bragg filtering for detecting material properties and discuss two
potential applications of this segmentation model (Section \ref{sec:results}). 
Furthermore, we show how it can be easily used to further automate software
based scientific image analysis pipelines. Finally, we summarize results and
suggest future extensions and uses (Section \ref{sec:conclusions}).

\section{Materials and Methods}
\label{sec:materail}




We first describe the process through which we grow our graphene samples, how images are extracted, and the different classes of surface structures observed in the data. We then introduce the mathematical definitions for the preprocessing of this acquired data used in this study. We then describe the conventional Bragg filtering method for segmentation. Finally we introduce our deep learning approach to the segmentation task.

\subsubsection{HRTEM Imaging of Graphene Structures}

The single-layer, polycrystalline graphene samples are grown on polycrystalline copper substrates at 135~$^\circ$C by chemical vapor deposition. The copper substrate is first held under 150 mTorr of pressure in hydrogen for 1.5 hours, and then 400 mTorr pressure of methane is flowed at 5 standard cubic centimeters per minute (sccm) to form single-layer graphene.  Further information of this sample preparation method are given by \cite{li2009large, rasool2011continuity, rasool2013measurement}.

\begin{figure}[htbp]
    \centering
    \includegraphics[width=3 in]{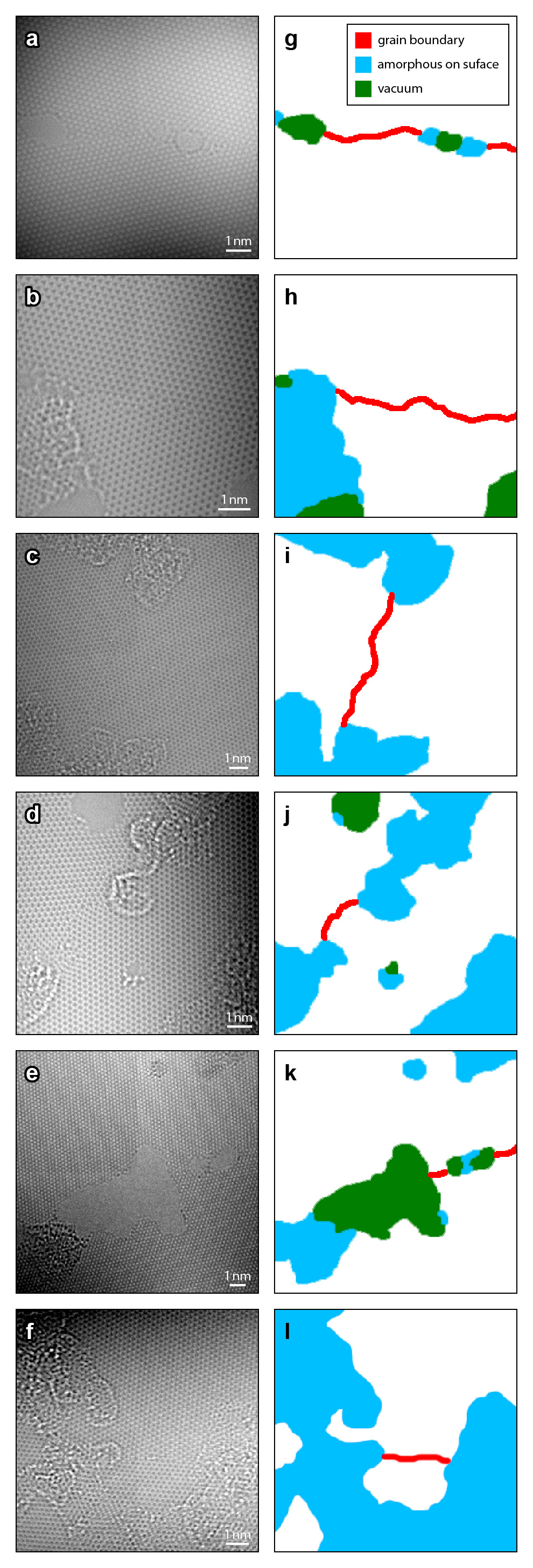}
    \caption{(a)-(f) HRTEM image examples of polycrystalline, single layer graphene, and (g)-(l) the corresponding labels.}
\label{figure:HRTEM_graphene}
\end{figure}

The majority of graphene HRTEM images utilized in the current study are published along with the measured atomic coordinates in a previous study \citep{ophus2015large}. Some additional HRTEM images, including those from time- and focal-series are from various studies of the structure of graphene grain boundaries \citep{rasool2013measurement, rasool2014conserved, ophus2017automatic} were also included in the image dataset. All of our HRTEM images of graphene were recorded on the TEAM 0.5 microscope, a monochromated and aberration-corrected FEI / Thermo Fisher Titan microscope operated at 80 kV. The imaging conditions are optimized for fast data collection with a relatively low electron dose in order to record as many images as possible. The dose varied from approximately 1~000 to 10~000 electrons / $\rm{\AA}$ngstrom$^2$ across all images.



Figs.~\ref{figure:HRTEM_graphene}a-f show examples of these HRTEM images.  Note that due to the monochromation of the electron beam, the intensity of all images varies across the field of view. The graphene samples used in this study contain four primary structural classes. The first class is the graphene lattice itself, which consists of a periodically repeating honeycomb structure. In this structure, each carbon atom is bonded to 3 neighbors with $120^\circ$ between each bond, and 6 carbon atoms form closed hexagonal rings, which are tiled in a close-packed triangular lattice. Figs.~\ref{figure:HRTEM_graphene}g-l show these regions marked in white. Depending on the microscope defocus, the contrast is either white-atom or black-atom, meaning either increased or decreased intensity at the location of each carbon atom respectively \citep{okeefe2008seeing}. Figs.~\ref{figure:HRTEM_graphene}b, c, d, and f show examples of white-atom contrast, while Figs.~\ref{figure:HRTEM_graphene}a and e show black-atom contrast. These images show varying degrees of residual imaging aberrations; this is a consequence of the low-dose measurement protocol used where the sample is exposed to as little electron fluence as possible. 

We can define a simple order parameter calculated by using image convolution to measure the difference in signals between the atoms on a hexagonal ring (using the measured graphene lattice parameter) and the center of the ring.  If we calculate this parameter for a range of hexagon orientations, the signal reaches a maximum when the measurement is oriented the same as the underlying lattice, giving an estimate for the lattice orientation. The regions where two different orientations meet in a discontinuity define the second class, the graphene grain boundaries. These boundaries are shown as red lines in Figs.~\ref{figure:HRTEM_graphene}g-l, and were the focus of the previous study by \cite{ophus2015large}.




The third class is the vacuum regions, marked as green areas in Figs.~\ref{figure:HRTEM_graphene}g-l. No structure is present in these regions, and the electron beam passes straight through with no modulation. Each of these first three classes is easy to detect with simple algorithms.  Measuring the location of the vacuum regions is trivial after illumination flat-field correction (described below) since these regions have unit intensity everywhere. 

However the fourth class, which corresponds to surface contaminants such as amorphous carbon, is more difficult to accurately segment. These regions are shown as blue areas in Figs.~\ref{figure:HRTEM_graphene}g-l. The regions often show strong lattice contrast of ideal or near-ideal graphene structure, overlaid with random modulations. These modulations can be strong or weak, and consist of a complex mix of white-atom and black-atom contrast. Amorphous contaminants also tend to be attracted to the structures we would like to analyze, e.g. the grain boundaries. This is likely due to the surface topology induced by these boundaries \citep{ni2019topological}. In the study of graphene grain boundaries by \cite{ophus2015large}, most of the analysis steps were automated, but avoiding these surface contaminant regions was done manually. In this work, we tackle the segmentation of these regions, as it represents the most difficult step to automate.



\subsubsection{Image Preprocessing}

For both the U-Net and Bragg filtering image segmentation, we have applied the same pre-processing and normalization steps, based on the image processing described in \cite{ophus2015large}. To normalize the intensity variation due to the monochromation, we have fit the average local intensity $I_0(\mathbf{r})$ for each image with a 2x2 B\'ezier surface \citep{Farin2001} given by the Equation \ref{eqn:invar}.

\begin{eqnarray}
\label{eqn:invar}
    I_0(\mathbf{r}) 
    &=& 
        k_{i,j}
        \sum_{i=0}^m
        {m \choose i} u^i (1-u)^{m-i}
        \nonumber \\
    && \hspace{12 pt} \cdot 
        \sum_{j=0}^n
        {n \choose j} v^j (1-v)^{n-j},
\end{eqnarray}
where $(u,v)$ are the image coordinates normalized to range from 0 to 1, and $k_{i,j}$ are the B\'ezier surface coefficients. After fitting these coefficients, the normalized intensity $I(\mathbf{r})$ is given by the Equation \ref{eqn:inten}.

\begin{equation}
\label{eqn:inten}
    I(\mathbf{r}) = 
    \frac{I_{\rm{meas}}(\mathbf{r}) }{I_0(\mathbf{r})},
\end{equation}
where $\mathbf{r}=(x,y)$ are the real space coordinates, and $I_{\rm{meas}}(\mathbf{r})$ is the measured image intensity. After this step, the mean intensity is equal to one. 

Next, we scale the intensity range by calculating the image standard deviation $\sigma$, equal to the root-mean-square of the intensity $\sigma = \sqrt{\langle (I(\mathbf{r})-1)^2 \rangle}$. We then normalize the image by subtracting the mean $\mu$ and dividing by the standard deviation $\sigma$ as described by Equation \ref{eqn:output}.


\begin{equation}
\label{eqn:output}
    I_{\rm{output}} =
    \frac{I(\mathbf{r}) - \mu}
    {\sigma} 
\end{equation}

The images in this dataset were originally 1024x1024 or 2048x2048 images. We resize these images down to the size of 256x256 pixels. Training/test labels were generated by hand using the Paint S software application for macOS. 

\subsubsection{Segmentation by Fourier Filtering}

\begin{figure*}[htbp]
    \centering
    \includegraphics[width=5.5 in]{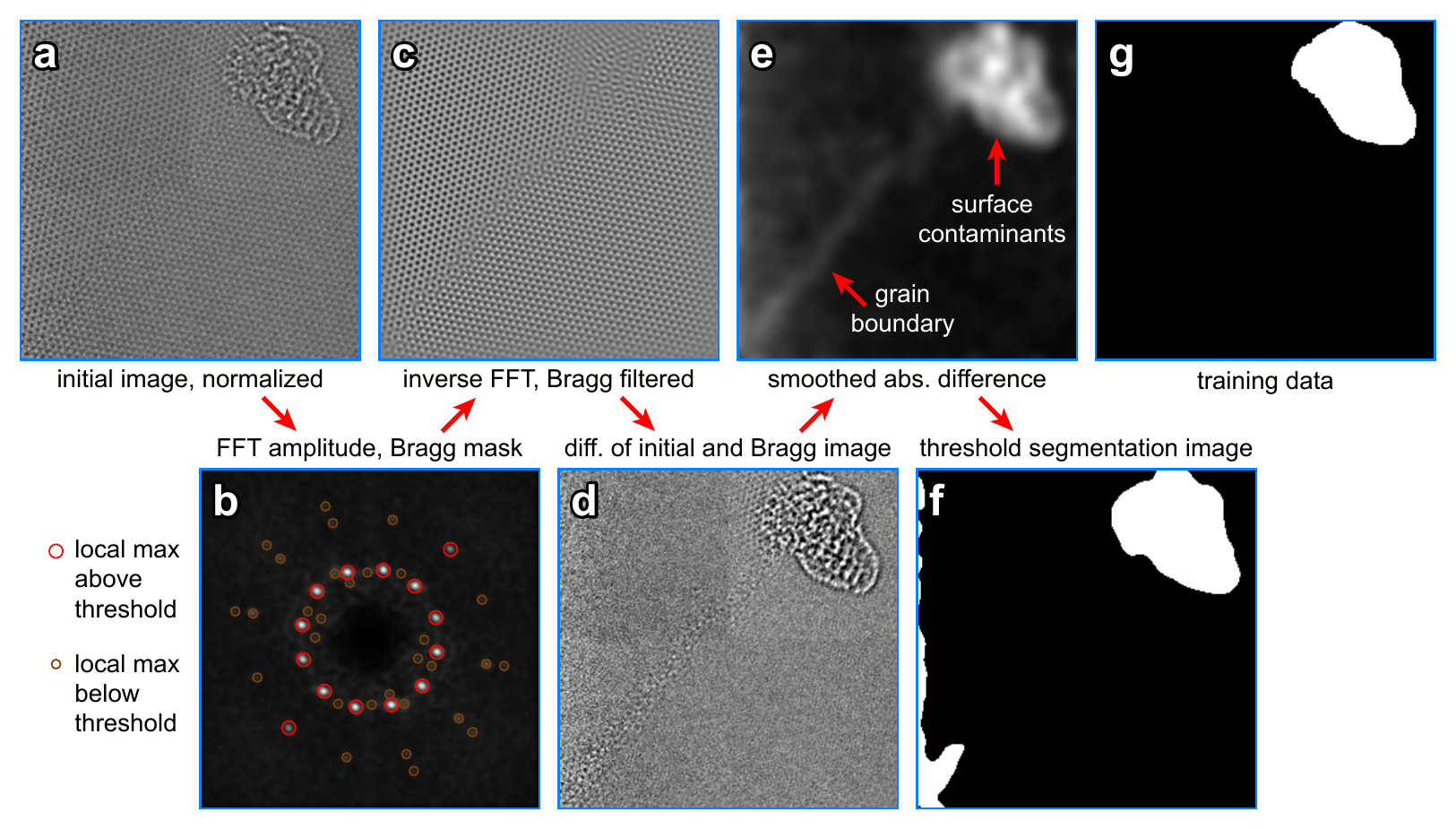}
    \caption{Segmenting surface contaminants and graphene lattices using Bragg filtering. (a) Input image data and (b) its Fourier transform amplitude. (c) Inverse Fourier transform after applying Bragg mask, and (d) difference from input image data. (e) Absolute difference smoothed and then (f) threshold segmentation image. (g) Corresponding training data.}
\label{figure:Bragg}
\end{figure*}

The defining feature for crystalline samples is their high degree of ordering and long-range translation symmetry. When crystalline samples are imaged along a low index zone axis, the resulting images display local periodicity inside each crystalline grain. These periodic regions create sharply peaked maxima in the image's 2D Fourier transform amplitude that are closely related to Bragg diffraction from a periodic crystal. These peaks are not strictly speaking due to true Bragg diffraction, but are nevertheless often referred to as ``Bragg spots''  \citep{hytch1997geometric, galindo2007peak}. By applying numerical masks around a given Bragg spot, we measure the degree of local ordering over the image coordinates that corresponds to the associated crystal planes \citep{pan1998quantitative}. We have developed a ``Bragg Filtering'' procedure to segment the images into two classes, corresponding to clean atomically resolved graphene, and the amorphous surface contaminants. Bragg filtering is a standard procedure in many image processing routines for atomic resolution micrographs such as lattice strain deformation mapping \citep{hytch1997geometric}. Our segmentation procedure is shown schematically in Fig.~\ref{figure:Bragg}. 

After preprocessing the initial image and padding the boundaries, we calculate a weighted Fourier transform $G(\mathbf{q})$ of the image $I(\mathbf{r})$ shown in Fig.~\ref{figure:Bragg}a, given by the Equation \ref{eqn:bragg}.

\begin{equation}
\label{eqn:bragg}
    G(\mathbf{q}) = |\mathbf{q}|
    \,
    | \mathcal{F}_{\mathbf{r} \to \mathbf{q}}
    \left\{
    I(\mathbf{r}) W(\mathbf{r})
    \right\}|,
\end{equation}
where $\mathbf{r}=(x,y)$ and $\mathbf{q}=(q_x,q_y)$ are the real space and Fourier space coordinates respectively, $\mathcal{F}_{\mathbf{r} \to \mathbf{q}}$ is a 2D Fourier transform from real to Fourier space, and $W(\mathbf{r})$ is a window function. Next, we find the local maxima of this image that are above a threshold value $G_{\rm{thresh}}$, shown in Fig.~\ref{figure:Bragg}b.

Next, we perform Bragg filtering by applying a 2D Gaussian distribution aperture to $N$ Bragg peaks at positions $\mathbf{q}_n$, given by the Equation \ref{eqn:braggfil}, where $\mathcal{F}_{\mathbf{q} \to \mathbf{r}}$ is the inverse Fourier transform, and $\sigma$ is the aperture size of the Bragg filter.

\begin{equation}
\label{eqn:braggfil}
    I_{\rm{Bragg}}(\mathbf{r}) = 
    \mathcal{F}_{\mathbf{q} \to \mathbf{r}}
    \left\{
    \mathcal{F}_{\mathbf{r} \to \mathbf{q}}
    \left\{
    I(\mathbf{r})
    \right\}
    \sum_{n=1}^N
    e^{ 
    -|\mathbf{q} - \mathbf{q}_n|^2 / 2 \sigma^2
    } \right\},
\end{equation}

Note that if symmetric pairs of Bragg diffraction peaks are used, the output image will be real-valued for all pixels. The resulting image is shown in Fig.~\ref{figure:Bragg}c. By subtracting the Bragg filtered image and the mean intensity from the original image, we generate an image consisting of the non-periodic components, shown in Fig.~\ref{figure:Bragg}d. Since we consider both negative and positive deviations to be signals from the surface contaminants, we take the absolute value and then low pass filter $F_{\rm{LP}}( ... )$ the output, giving an image like Fig.~\ref{figure:Bragg}e. In this figure, we see weak signals in the aperiodic boundary between the two graphene grains, and a strong signal from the contaminants. 

Finally, by choosing an appropriate mask threshold $M_{\rm{thresh}}$, we compute the desired segmentation output $I_{\rm{seg}}(\mathbf{r})$ by the Equation \ref{eqn:output}. The output is shown in Fig.~\ref{figure:Bragg}f.

\begin{equation}
    I_{\rm{seg}}(\mathbf{r}) =
         F_{\rm{LP}} \Big[
            | I(\mathbf{r})  -
            I_{\rm{Bragg}}(\mathbf{r}) |
         \Big] 
         > M_{\rm{thresh}}
\end{equation}

This image compares favorably to the training dataset in Fig.~\ref{figure:Bragg}g. We accurately mask the contaminant region, while not producing a ``false postive'' signal at the grain boundary. There are some false positives  at the image boundary due to the breakdown of the lattice periodicity at the image boundaries. We use padding and normalization of the filter output to reduce the magnitude of these effects, but they are still present in some images. In practice however, these edge artifacts are insignificant, since we cannot perform accurate measurements of the local atomic neighborhood at the image boundaries.


In this study, we have optimized the parameters $\sigma$, $G_{\rm{thresh}}$, and $M_{\rm{thresh}}$ by using gradient descent to minimize the error between the segmentation maps generated and the training data. The Bragg filter parameters and performance was measured using five-fold cross validation, giving values of $\sigma = 0.0156 \pm 0.0005$ 1/pixels, $G_{\rm{thresh}} = 12.5 \pm 0.1$, and $M_{\rm{thresh}} = 0.054 \pm 0.000$. The maximum number of Bragg peaks included was coarsely tuned, but does not strongly affect the results (as long as it is high enough) and is therefore fixed at 24. The degree of low pass filtering both of the initial Fourier transform for Bragg peak detection, and of the output difference image does not strongly affect the results. These two steps were performed by convolution with a 2D Gaussian function with 2 and 5 pixels standard deviation respectively. The cross-validation ensures that the segmentation map accuracy was measured only on the validation subset of the data, using parameters optimized from the other 80\% of the training images. This procedure prevents  over-fitting of the parameters to the training data. The accuracy of the resulting segmentation images is summarized in Table \ref{tab:performance_table}.


 

\subsubsection{Deep Learning Segmentation}

The U-Net architecture \citep{unet} is a CNN architecture based on a fully convolutional neural network, modified and extended to improve segmentation performance for medical and scientific (microscopy) segmentation tasks with limited training data. A fully convolutional neural network (FCN) is a common baseline deep learning architecture for semantic segmentation. It is formed using convolutional layers, pooling layers, nonlinear activation layers, and transposed convolution or upconvolution layers. It generates a network output equal in dimensions to the input, offering a classification for each input pixel. U-Net, which is a network based on the FCN, consists of a down-sampling path and an up-sampling path. 
Features from the down-sampling path are combined with features generated by the up-sampling path.

Each U-Net block in the down-sampling path consists of two $3\times 3$ kernel convolution layers, a Rectified Linear Unit (ReLU) layer and a a $2\times 2$ Maximum Pooling operation (MaxPool) layer. Each block in this path doubles the number of features in the previous block. Each block in the up-sampling path starts with an up-convolution operation that doubles the size of the feature maps but also reduces the number of feature by a factor of two. The features from this up-convolution operation are concatenated with the feature maps of the corresponding layer of the down-sampling path, bridged across the network. This is followed by two convolution layers, and a single ReLU layer. The final layer in each U-Net block is a single $1\times 1$ convolution layer that translates the final feature maps to two separate output classes, foreground and background.

\begin{figure}[htbp]
    \centering
    \includegraphics[width=3.25in]{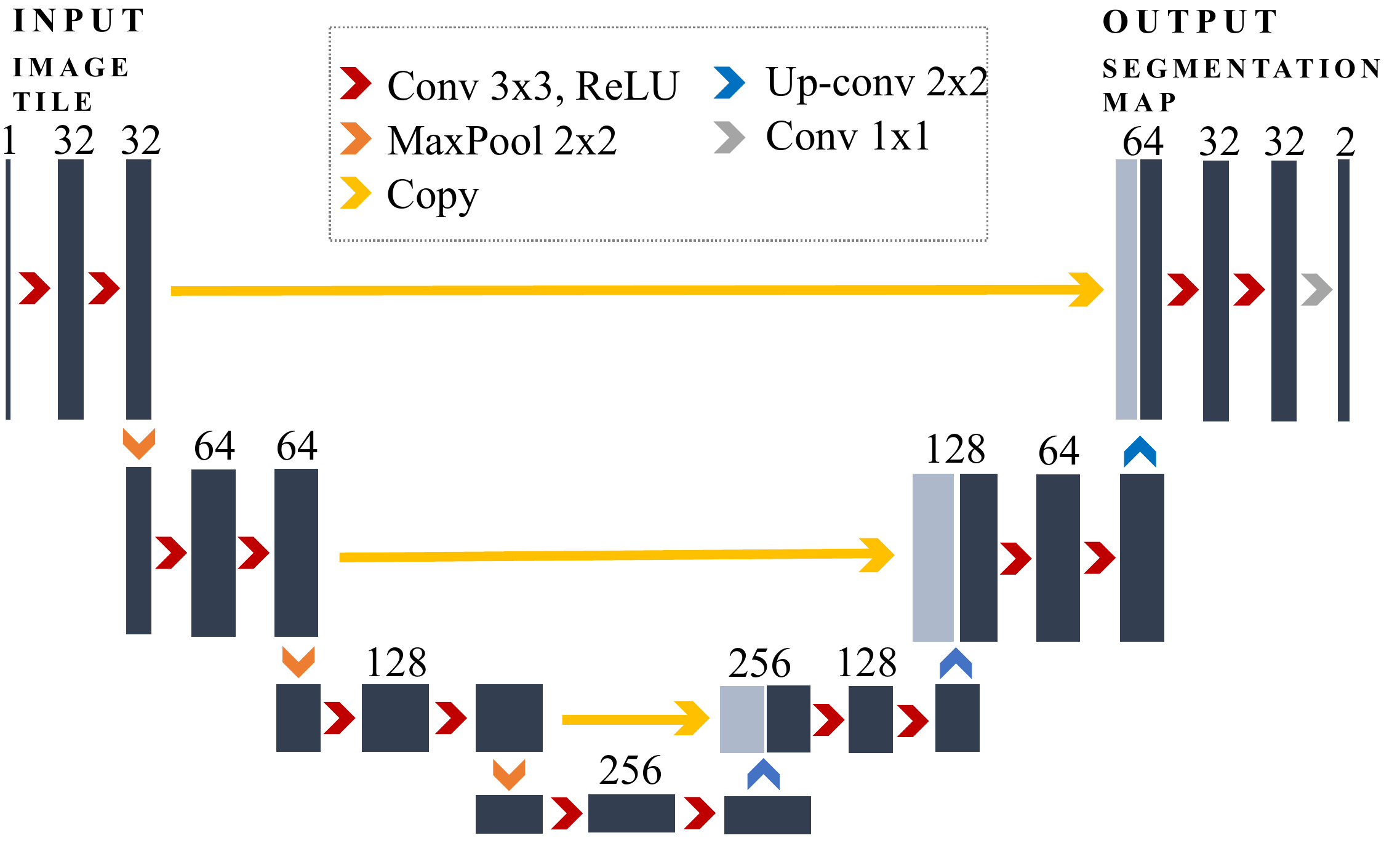}
    \caption{Our Neural Network Architecture based on a scaled down version of U-Net. Each color arrow corresponds to the different operation: 2D Convolution (Conv) with the kernel size of 3x3, Rectified Linear Unit (ReLU), Maximum Pooling (MaxPool) with the window size of 2x2, feature map Copying (Copy), 2x2 Up (Up-conv). Each dark grey box represents a multi-channel feature map resulting from the previous convolution operation, where light grey boxes represent features copied from the down path to the up path. The number of feature maps between each convolution layer is labeled at that top of the box.}
\label{figure:architecture}
\end{figure}

The original U-Net architecture consists of a total of 9 U-Net blocks.  
The modified U-Net architecture used in our work is illustrated in Fig.~\ref{figure:architecture}. First, we use padded convolution, allowing us to produce a segmentation map with the same size as the input image. In order to build a model optimized to our data, we need to select the best hyperparameters.  In the context of deep learning, hyperparameters are configurations set by the developer for a given data modeling problem. These are typically factors such as the learning rate, number of features, and depth of the network. The optimal hyperparameters for a given data modeling problem are not known from the start and are usually optimized through trial and error. In this study, we used the grid search method for hyperparameter tuning. Grid search involves exhaustive trial and error of all combinations of possible hyperparameters in a search space defined by the programmer. Once each version of the network was trained for 150 epochs, the model with the best performance based on the accuracy score was selected. The selected model was then retrained for an additional 300 epochs for two trials to generate accuracy metrics in order to verify full convergence. We optimized across 3 hyperparameters: number of features in the first block (with each subsequent block having an adjusted number of features to maintain network symmetry), number of blocks, and learning rate. We found that a U-Net with 7 blocks and 32 features in the first block with a learning rate of .0001 achieved the highest accuracy of  .9792 and a Jaccard score of .8125. For reference, the original U-Net model achieved an accuracy score of .9775 and a Jaccard score of .8015 using a .0001 learning rate, which yielded the best results for this network architecture as well. An additional advantage of a smaller network is achieving higher throughput which is essential for speed dependant applications such as compression of real time tracking. 

We train this network using an Adam optimizer \cite{kingma2014adam} to estimate parameters and pixel-wise cross entropy as a loss function. We train for 300 epochs/approximately 41 minutes on an Nvidia an GeForce GTX TITAN Black GPU, using a learning rate of $0.0001$. We use 5 fold cross validation for training. In this scheme, the dataset is split into 5 different groups. For each unique group, the data in that group is excluded, and the model is fit (from scratch) to the remaining 4 groups. Then this model is evaluated for accuracy metrics on the excluded fold in order to have a separate test set from the training set. at the end, the accuracy metrics from the 5 folds are averaged together. 
We repeat this experiment twice over the different folds, randomizing the dataset each trial, and then average the performance results for evaluation. We train our model using the PyTorch Deep Learning Framework \citep{paszke2019pytorch}.

\section{Results and Discussion}

\label{sec:results}
In this section, we describe the results of our experiments. We first qualitatively compare the two methods in terms of the types of errors perceived via two separate cases. We then define several key performance metrics that are commonly used for evaluating segmentation models, and report the performance of each model across these different metrics. We also describe how these networks can be used in atomic position workflows and discuss the different results of using each model. Finally, we offer an additional usage of the segmentation model for data reduction and sorting.

\subsubsection{Qualitative Comparison}

\begin{figure*}[htbp]
    \centering
    \includegraphics[width=7 in]{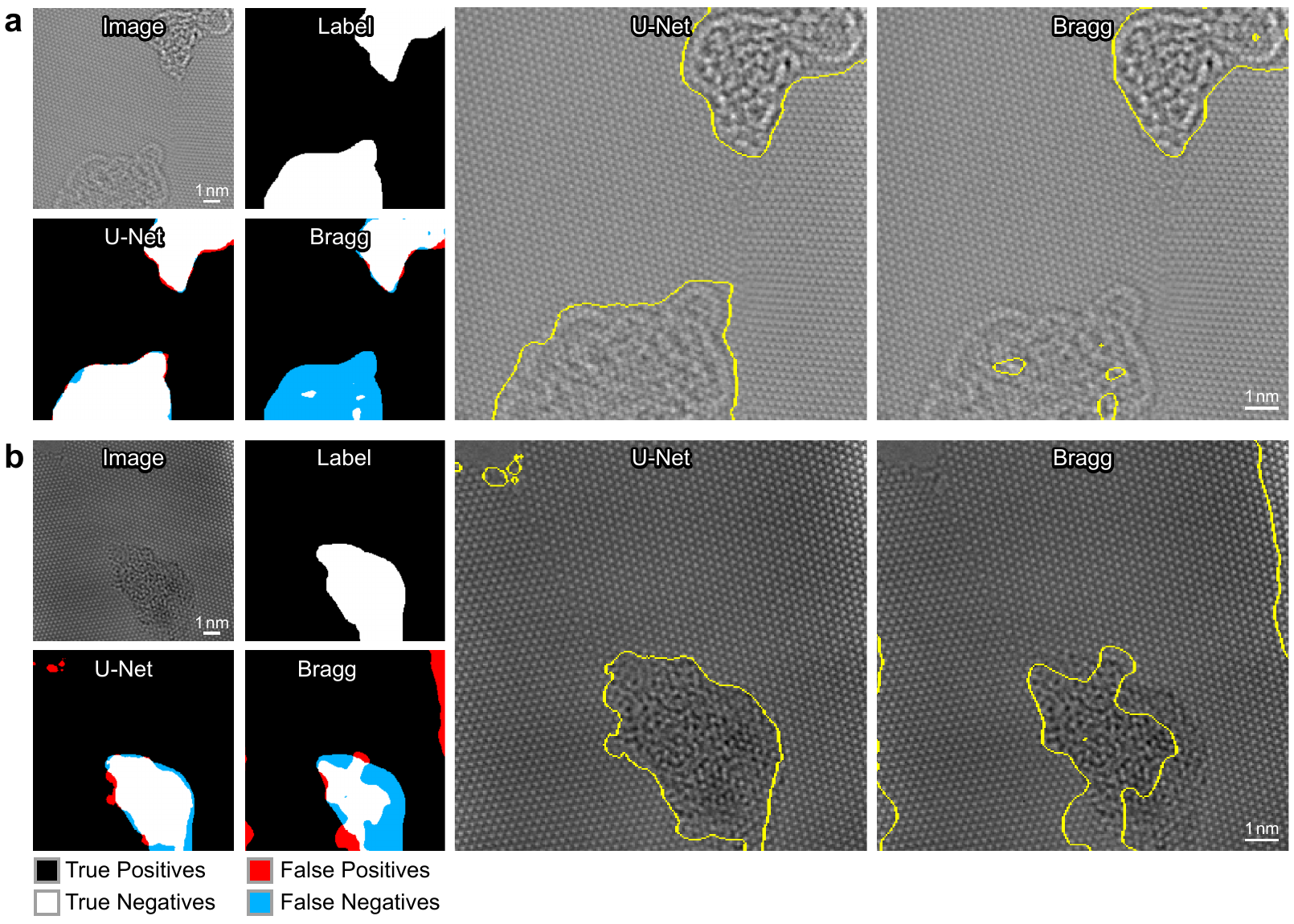}
        
    \caption{Comparison of the performance of the Bragg filter method and the U-Net method for segmentation. Detected errors are visualized on the left. True Positives are shown in white, True Negatives in black, false positives in red and false negatives in blue. The sectioning generated by U-Net is shown on the original image in the middle column, and the sectioning generated by the Bragg filter is shown in the right column. }
\label{figure:positive_negatives}
\end{figure*}

Fig.~\ref{figure:positive_negatives} shows the performance of each method on two different example input images. Fig.~\ref{figure:positive_negatives}a shows an example of a critical failure by the Bragg filter to detect amorphous carbon. The Bragg filter produces incomplete segmentation results ignoring the bottom section of amorphous carbon, possibly due to its low contrast with respect to the other region of amorphous carbon at the top of the image. On the other hand, U-Net successfully identifies both of these regions as amorphous with precision. The resolution of our segmented boundaries (positional uncertainty of the edges) is roughly given by the feature size of the structures being imaged, approximately equal to the graphene lattice parameter.

Fig.~\ref{figure:positive_negatives}b shows an example where both the U-Net and Bragg Filter methods produce precise results; however, the results of the Bragg filter produces 
a substantial amount of false negatives, where as the U-Net method produces visibly superior results. Furthermore, the bright spot at the right side of the image is incorrectly identified as deformed graphene. U-Net, in contrast, has a much higher coverage of the main defect present in the image, and does not incorrectly classify the bright spot as amorphous. Moreover, the U-Net segmentation detects false positives of amorphous graphene in the upper left hand corner. However, if one inspects the raw image, one will observe that there is in fact a small amount of amorphous graphene present in this part of the image that is overlooked during the labeling process. This demonstrates that the capabilities of deep learning surpass human performance on some tasks and effectively avoid over-fitting to human labels in certain cases.

\subsubsection{Quantitative Comparison}

\begin{table}[h!]
  \begin{center}
    \begin{tabular}{l r r r} 
      \hline
      {\bf Performance Metric} &
      {\bf U-Net} &
      {\bf Bragg} &
      {\bf Difference [\%]}\\
      \hline
      \hline
      
        Accuracy Score &  \textbf{0.979} &   0.956  &    2.4\%\\
        \hline
        Balanced Accuracy Score &  \textbf{0.950} &   0.897 &    5.3\%\\
        \hline
        Jaccard Score &  \textbf{0.812} &  0.688 &   12.4\%\\
        \hline
        F1 Score &  \textbf{0.884} &  0.791  &   9.2\% \\
        \hline
        Precision Score &  \textbf{0.886} &  0.829  & 5.6\% \\
        \hline
        Recall Score &  \textbf{0.889} &  0.792 & 9.6\%  \\
        \hline
              
    \end{tabular}
    \caption{Table of Performance Metrics}
    \label{tab:performance_table}
  \end{center}
\end{table}

\begin{table}
  \begin{center}
    
    \begin{tabular}{l|c|r} 
      
      $ $ & $Y'=1$ & $Y'=0$ \\
      \hline
      $Y=1$ & TP & FN \\
      $Y=0$ & FP & TN \\
      
    \end{tabular}
    \caption{Definitions of True Positive, False Positive, True Negative, and False Negative, where Y is the ground truth and Y' is the value predicted by a model.}
    \label{tab:TF_table}
  \end{center}
\end{table}

Multiple metrics for evaluting segmentations and comparing models exist~\citep{metrics, jayaram2002methodology}, where the
choice of a particular metric depends on the problem scope and goals. In the following we consider and define several
metrics, discuss their typical usage as well as limitations, and use them to compare our U-Net-based approach to Bragg filtering.
Table~\ref{tab:performance_table} summarizes all results, which were computed using SciKit-Learn~\citep{pedregosa2011scikit}. All metrics
are defined in terms of true positives (TP), true negatives (TN), false positives (FP), and false negatives (FN), see Table~\ref{tab:TF_table}.

One typical score for evaluating a segmentation is \textsl{pixel wise accuracy}, calculated as the total number
of correctly classified pixels (TP, TF) divided by the number of predictions (i.e., total number of pixels) 
\begin{displaymath}
\text{PixelwiseAccuracy} = \frac{TP+TN}{TP+TN+FP+FN}\quad\text{.}
\end{displaymath}
Both segmentation methods achieve a pixel wise accuracy above 95\%, with U-Net performing roughly 2\% better.
While this metric is easy to interpret, it fails to properly take large class imbalances into account.
In our case, segmentations are dominated by the background class, making the results of this metric
potentially skewed. To avoid performance metric inflation due to class imbalance, \textit{balanced accuracy} weighs samples by the
inverse prevalence of their true classes. For the binary case, balanced accuracy is computed as
\begin{displaymath}
\text{BalancedAccuracy} = \frac{1}{2} \left( \frac{TP}{TP+FN}+\frac{TN}{TN+FP} \right)\quad\text{.}
\end{displaymath}
Using this metric, we start to see the performance of our models to diverge numerically,
with a difference of around 5\% in favor of U-Net.

The \textit{Jaccard score} and the \textit{F1 score} are two of the most common metrics used for evaluating
semantic segmentation models or in the presence of class imbalances. Mathematically, both metrics evaluate the
same aspects of model performance \citep{metrics}.
The Jaccard score is computed as 
\begin{displaymath}
\rm{Jaccard} Score = \frac{TP}{TP+FP+FN}\quad\text{,}
\end{displaymath}
and the F1 score is computed as
\begin{displaymath}
\rm{F1 Score} = \frac{2TP}{2TP+FP+FN} \quad\text{.}
\end{displaymath}
Our U-Net model shows Jaccard and F1 improvements over the Bragg filtering method of approximately 12\% and 9\% respectively. This further shows that for our class imbalanced dataset, U-Net holds significant advantages over Bragg filtering that are not as clear when using other metrics like accuracy. 

\begin{figure*}[htbp]
    \centering
    \includegraphics[width=6.5 in]{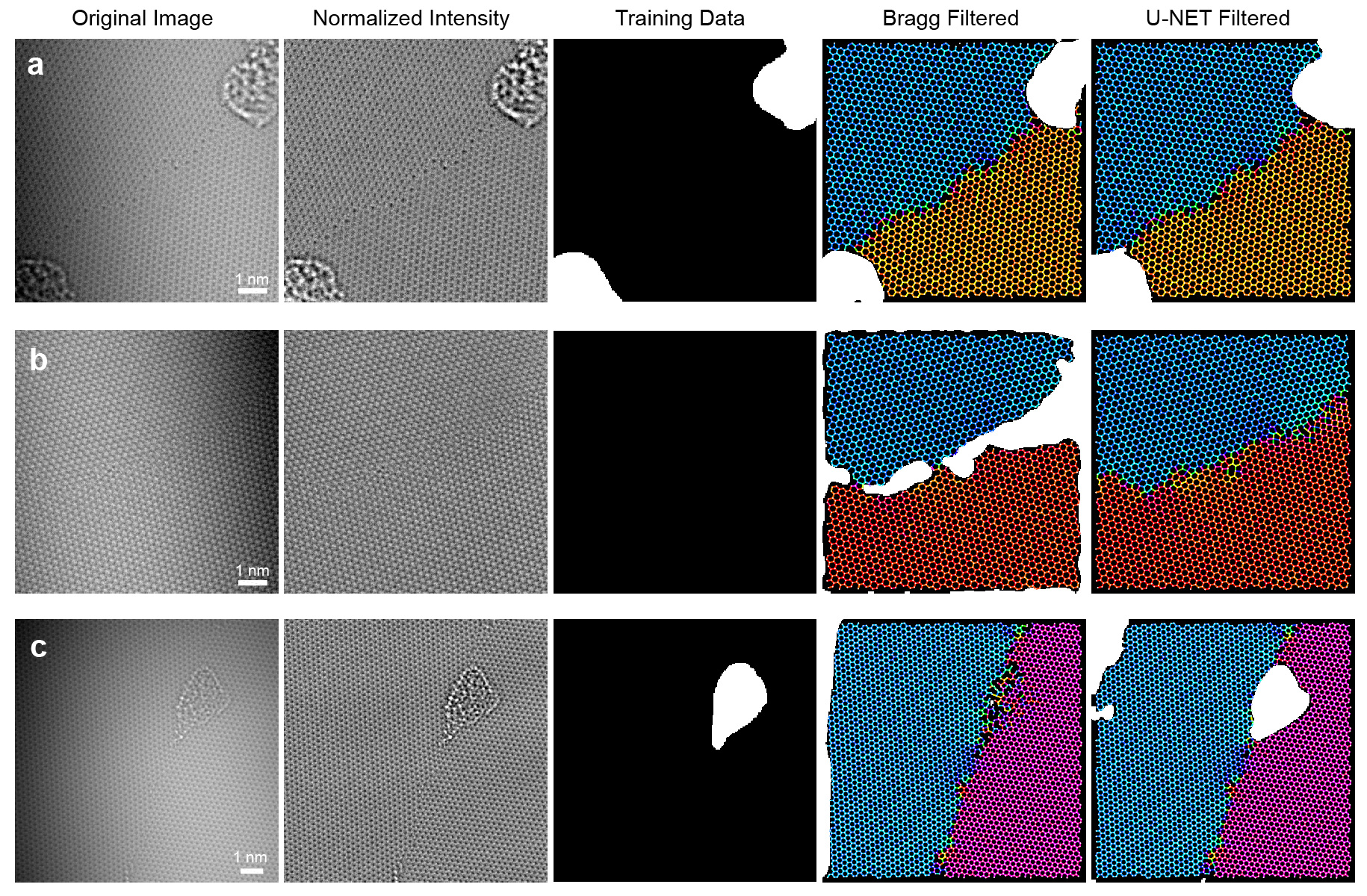}
    \caption{Three examples comparing Bragg and U-net filtering. (a) Output masks where both filters performed well, (b) masks where Bragg filter produced false positive regions, and (c) masks where Bragg filter produced false negative regions. In both (b) and (c), the U-Net filter produced a more accurate result.}
\label{figure:atomic_workflow}
\end{figure*}

To analyze performance differences more specifically in terms of in terms of false positives and false negatives, we compare
the \textit{precision} and \textit{recall scores}. \textit{Precision} represents the ratio of correctly predicted positives
to total predicted positives, i.e.,
\begin{displaymath}
\rm{Precision} = \frac{TP}{TP+FP} \quad\text{,}
\end{displaymath}
while \textit{recall} is the ratio of correctly predicted positives to the total number of positives in the ground truth, i.e.,
\begin{displaymath}
\rm{Recall} = \frac{TP}{TP+FN} \quad\text{.}
\end{displaymath}
 High precision corresponds to a low false positive rate, where as high recall
 indicates to a low false negative rate. Our study shows that U-Net achieves
 around 4\% and 10\% improvement over the Bragg filtering method in precision and
 recall scores respectively.
 
 \subsubsection{Memory Usage and Speed}
 
 Using a workstation running an Intel Xeon 6138 CPU, the Bragg filtering routine requires 0.0203 seconds to process each 256x256 image. The RAM requirements for this routine are approximately 10 times those required for a single image, equal to 5 MB for a 256x256 pixel image. The U-Net implementation requires 0.0294 seconds per image, and uses 710MiB GPU RAM using an NVIDIA Titan Black GPU, for 256x256 size images.

\subsubsection{Segmentation in Atomic Position Workflows}


First, we demonstrate the use of Bragg and U-Net segmentation as part of an automated
atom and bond finding routine (Fig.~\ref{figure:atomic_workflow}). After an initial
illumination correction step, we determine candidate atom positions using the method
described by \cite{ophus2015large}. Subsequently, we discard all atom positions that
fall inside contamination regions, determined either by using the mask output of the
Bragg filtering routine or the U-Net segmentation routines, and remove them from the
list of candidates. Finally, we connect neighboring atoms within a given distance
threshold by candidate atomic bonds. Fig.~\ref{figure:atomic_workflow} shows
candidate atom positions as white dots and candidate bonds colored by the bond angle modulo 60$^\circ$,
overlaid on the masks generated by the two segmentation filters.
For this paper, we do not perform further any refinement of the atom positions and bonds
and focus on segmentation quality.

Fig.~\ref{figure:atomic_workflow}a shows scenario where both the Bragg filter and U-Net produce suitable results.
In both cases, two surface contamination regions are correctly identified and masked, in agreement with the training data.
In contrast, Fig.~\ref{figure:atomic_workflow}b shows an example where the Bragg filter produces a series of false positive regions (type I errors) at the boundary between the two graphene grains. This error can be addressed by decreasing the masking threshold $M_{\rm{thresh}}$, but this change would overall increase the error across the full dataset. Fig.~\ref{figure:atomic_workflow}c shows an example of the kind of errors introduced into the Bragg filter segmentation when $M_{\rm{thresh}}$ is set too low. In this example, the Bragg filter failed to mask off a region of surface contamination, which in turn leads to many erroneous atom positions and bonds. These two examples illustrate one weakness of the conventional Bragg filtering routine, which is that it relies on a small number of hyperparameters that cannot be set to values that will successfully perform the segmentation across the full dataset.

In comparison, the U-Net segmentation shown in Fig.~\ref{figure:atomic_workflow}b and c outperforms the Bragg filter and produces successful results. These examples show that the deep learning approaches can be both more accurate and more robust than simpler, conventional imaging filters, i.e. Bragg filter. Both segmentation methods produce a false positive at the left edge of Fig.~\ref{figure:atomic_workflow}c, due to the correction of the low initial intensity value in that region causing boosting of the noise. However, edge pixels are significantly less valuable in analysis of the atomic structure because the neighboring atomic environment is not visible at image boundaries. 

\subsubsection{Data Reduction and sorting}

\begin{figure*}[t!]
    \centering
    \includegraphics[width=6.3 in]{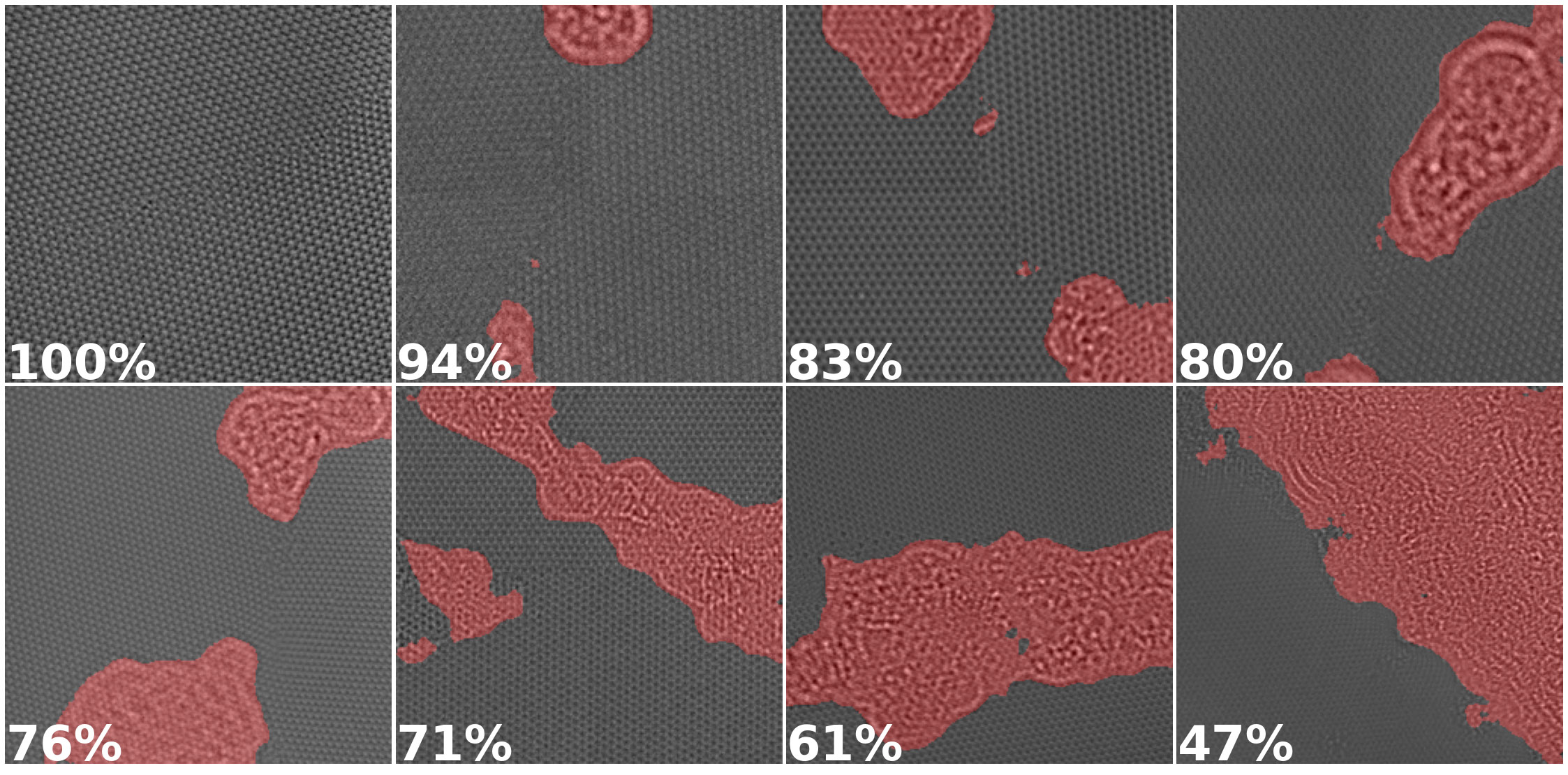}
    \caption{A subset of the images from our dataset sorted by the percentage of pixels that contain potentially useful information as computed by U-Net in decreasing order from left to right. Regions classified as amorphous carbon are highlighted in red.}
\label{figure:useful}
\end{figure*}

Lastly, we demonstrate how our segmentation methods can be applied to a data reduction and image sorting problem. Scientific experiments can yield tens thousands of images which can often have a high variance in terms of information that is useful for the particular experiment being practiced. To be able to sort images by content ratio could be very useful to scientists. In the case of graphene images, a scientist may want to sort images by the ratio of graphene to amorphous carbon.
We demonstrate the capability of our segmentation models to perform this task in Fig.~\ref{figure:useful}. Images are sorted based on what percentage is dominated by graphene rather than amorphous carbon. 
In some cases where frame rates of the data acquisition instrument are too high for the transference of the data to be done in a reasonable amount of time, it might be useful to compress certain parts of the image data. One way approach to this is the implementation of a run length compression on the content class that may be less important to that specific experiment being carried out. 

\section{Conclusion}

In this study, we have compared two methods to perform segmentation of complex features in phase contrast HRTEM images of monolayer graphene. The two methods we used were a conventional Bragg filtering algorithm, and a deep learning method utilizing the U-Net architecture. The U-Net filter outperformed the conventional method in every performance metric tested, and was very robust against incorrect determination of structurally important regions. The U-Net method has the additional advantage of being adaptable to many different pixel-wise classification problems, and only requires a labeled dataset with a sufficient number of images that contain the desired segmentation features  to perform the training. In the future, it may be possible to randomly generate structures and perform image simulation to automatically generate labeled training datasets, removing even this relatively minor barrier. Because of their generality and robustness, deep learning methods such as U-Net segmentation are extremely valuable for fully-automated image processing in TEM.

\begin{acknowledgements}

\section*{Source Code Availability}

The adapted U-Net source code, HRTEM images and the amorphous region labeled images for training are all available at our 
\href{https://github.com/lbnlcomputerarch/graphene-u-net}{DL\_Segmentation\_HRTEM Github repo}

\section*{Acknowledgements}

This work was supported by the Office of Advanced Scientific Computing Research for the Computational Research Division, of the U.S. Department of Energy under Contract No. DE-AC02-05CH11231. Work at the Molecular Foundry was supported by the Office of Science, Office of Basic Energy Sciences, of the U.S. Department of Energy under Contract No. DE-AC02-05CH11231. RS, AB, and GHW acknowledge additional support of the Laboratory-Directed Research and Development project ``Network Computing for Experimental and Observational Data Management.'' CO acknowledges additional support from a Department of Energy Early Career Research Program award.

\end{acknowledgements}

\bibliography{main} 

\end{document}